\documentclass[journal, twocolumn, 10pt]{IEEEtran}
\usepackage{graphicx}
\usepackage{algorithm}
\usepackage{algorithmicx}
\usepackage{algpseudocode}
\usepackage{cite}
\usepackage{color}
\usepackage{amsmath,amssymb,amsfonts,amsthm,dsfont}
\usepackage{bm}
\usepackage{setspace}
\usepackage{multirow}
\usepackage{mathtools}
\usepackage{epstopdf}

\begin{document}
\title{Supervised Collective Classification for Crowdsourcing}
\author{
    \IEEEauthorblockN{Pin-Yu Chen\IEEEauthorrefmark{1}, Chia-Wei Lien\IEEEauthorrefmark{2}, Fu-Jen Chu\IEEEauthorrefmark{3}, Pai-Shun Ting\IEEEauthorrefmark{1},
Shin-Ming Cheng\IEEEauthorrefmark{4} } \\
    \IEEEauthorblockA{\IEEEauthorrefmark{1}Department of Electrical Engineering and Computer Science, University of Michigan, Ann Arbor, USA
    \\ \{pinyu, paishun\}@umich.edu} \\
    \IEEEauthorblockA{\IEEEauthorrefmark{2}Amazon Corporate LLC
    \\ chiaweil@amazon.com } \\
    \IEEEauthorblockA{\IEEEauthorrefmark{3}Institute for Robotics and Intelligent Machines, Georgia Institute of Technology, Atlanta, USA
    \\ fujenchu@gatech.edu} \\
    \IEEEauthorblockA{\IEEEauthorrefmark{4}Department of Computer Science and Information Engineering, National Taiwan University of Science and Technology, Taipei, Taiwan
    \\ \{smcheng\}@mail.ntust.edu.tw}
}

\maketitle
\setstretch{1.0}
\thispagestyle{empty}
\begin{abstract}
Crowdsourcing utilizes the wisdom of crowds for collective classification via information (e.g., labels of an item) provided by labelers. Current crowdsourcing algorithms are mainly unsupervised methods that are unaware of the quality of crowdsourced data. In this paper, we propose a supervised collective classification algorithm that aims to
identify reliable labelers from the training data (e.g., items with known labels). The reliability (i.e., weighting factor) of each labeler is determined via a saddle point algorithm. The results on several crowdsourced data show that supervised methods can achieve better classification accuracy than unsupervised methods, and our proposed method outperforms other algorithms.
\end{abstract}


\section{Introduction}
\begin{figure}[t]
    \centering
    \includegraphics[width=3.5in]{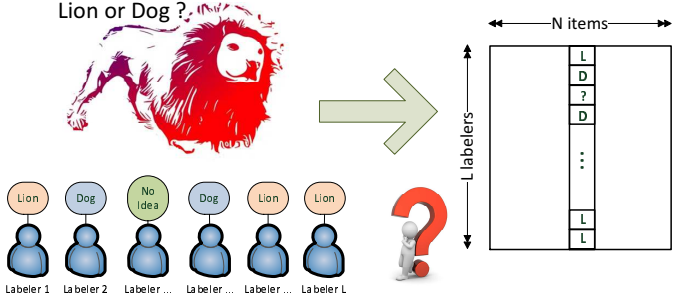}
    \caption{Illustration of collective classification for crowdsourcing.}
    \label{fig_LionDog}
\end{figure}

In recent years, collective decision making based on the wisdom of crowds has attracted great attention in different fields \cite{Surowiecki05}, particularly for
social networking empowered technology \cite{Robert11,Choffnes10,Robson12,Albors08} such as the trust-based social Internet of Things (IoT) paradigm  \cite{Nitti14,Chen15,Nitti15}.
Collective classification leverages the wisdom of crowds to perform machine learning tasks by acquiring multiple labels from crowds to infer groundtruth label.
For instance, websites such as \emph{Galaxy Zoo} asks visitors to help classify the shapes of galaxies, and \emph{Stardust@home} asks visitors to help detect interstellar dust particles in astronomical images. In addition, new business model based on crowdsourcing \cite{Howe06} has emerged in the past few years. For instance, \emph{Amazon Mechanical Turk (MTurk) }and \emph{CrowdFlower} provide crowdsourcing services with cheap prices. For \emph{MTurk}, a minimum of 0.01 US dollar is paid to a labeler/worker when she makes a click (i.e., generates a label) on an item. An illustrating figure can be found in Fig. \ref{fig_LionDog}.

Despite its cheap costs for acquiring labels, one eminent challenge for collective classification lies in dealing with these massive yet potentially incorrect labels provided by labelers. These incorrect labels may hinder the accuracy of collective classification when unsupervised collective classification methods (e.g., majority vote) are used for crowdsourcing.
Unsupervised collective classification using the expectation maximization (EM) algorithm \cite{Dempster77} is firstly proposed in \cite{Dawid79}. A refined EM algorithm is then proposed in \cite{Raykar10}, which is shown to outperform majority vote. In \cite{Zhou12}, a minimax entropy regularization approach is proposed to minimize the Kullback- Leibler (KL) divergence between the probability generating function of the observed data and the true labels. Some data selection heuristics are proposed to identify high-quality labels/labelers for collective classification based on weighted majority votes \cite{Dekel09,Ertekin11}.

By allowing a fairly small amount of items with known labels for collective classification,
it is shown in \cite{Ipeirotis10,Tang11,Zhou12,Ipeirotis14} that supervised collective classification can improve the classification accuracy within affordable costs.
Typical supervised classification algorithms include binary support vector machine \cite{Cortes95}, multi-class support vector machine \cite{Crammer02}, naive Bayes sampler \cite{Snow08,Tang11,Zhou12}, and multi-class Adaboost \cite{Zhu06}.

This paper provides an overview of the aforementioned methods and our goal is to propose a supervised collective classification algorithm that assigns weights to each labeler
based on the accuracy of their labels.
The weights are determined by solving a saddle point algorithm and they reflect the reliability of labelers. In addition to crowdsourcing,
the proposed method can be applied to other communication paradigms such as mobile sensing and cooperative wireless network by assigning more weights to reliable users.
For performance evaluation, we compare supervised and unsupervised collective classification algorithms on several crowdsourced datasets, which include a canonical benchmark dataset and the exam datasets that we collected from exam answers from junior high and high school students in Taiwan\footnote{The exam datasets are collected by the authors and  publicly available at the first author's personal website https://sites.google.com/site/pinyuchenpage}. The results show that the proposed method outperforms others in terms of classification accuracy.

\section{Problem Statement and Notations}
Consider there are $L$ labelers, $N$ items for classification, and $K$ label classes for items. Let $i \in \{1,2,\ldots,L\}$ denote the $i$-th labeler, $j \in\{1,2,\ldots,N\}$
denote the $j$-th item, and $x_{ij} \in\{0,1,2,\ldots,K\}$ denote the label of item $j$ given by labeler $i$. $x_{ij}=0$ if item $j$ is not labeled by labeler $i$.
For supervised data, each item is associated with a set of labels and a true label $\{X_j,y_j\}_{j=1}^{N_s}$, where $X_j=[x_{1j},x_{2j},\ldots,x_{Lj}]^T$ and $N_s$ is the number of supervised/training items.

For crowdsourced data, $X_j$ might be a sparse vector, where sparsity is defined as the number of nonzero entries in a vector. The sparsity originates from the fact that a labeler might only label a small portion of items. For the collected multiple choice exam data where labels are answers provided by students, $X_j$ is in general not a sparse vector. We aim to construct a classifier $f:\{0,1,\ldots,K\}^L \mapsto \{1,2,\ldots,K\}$ for collective classification based on supervised data.
For binary classification, we will use the convention that $x_{ij} \in \{-1,0,1\}$ and $y_j \in \{-1,1\}$. For multi-class classification, unless stated,
we use the one-to-all classifier $f(X_j)=\arg\max_{k\in\{1,2,\ldots,K\}} f_k(X_j)$, where $f_k(X_j)$ is the binary classifier of item $j$ such that the labels of class $k$ are $1$ and the labels of the other classes are $-1$. We also denote the predicted label of item $j$ by $\hat{y}_j$ and the indicator function by $\mathds{1}_{\{\cdot\}}$, where $\mathds{1}_{\{e\}}=1$ if event $e$ is true and $\mathds{1}_{\{e\}}=0$ otherwise. We further define the weight of each labeler by $w_i$ and define the weighting vector $w=[w_1,w_2,\ldots,w_L]^T$. For classifiers associated with weighting vector $w$, we have $f_k(X_j)=\textnormal{sign} (w^T X_j)$, where $\textnormal{sign}(z) =1$ if $z \geq 0$ and $\textnormal{sign}(z)<0$ otherwise.

\section{Overview of Crowdsourcing Algorithms}
\label{sec_existing_methods}

\subsection{Majority Votes (MV)}
Majority votes is the baseline (unsupervised) classifier for collective classification. The classifier is
\begin{align}
\label{eqn_majority_vote}
f^{MV}(X_j)={\arg\max}_{k\in\{1,2,\ldots,K\}} \sum_{i=1}^L \mathds{1}_{\{x_{ij}=k\}}.
\end{align}

\subsection{Weighted Averaging (WA)}
\label{subsec_WA}
Weighted averaging is a heuristic approach for weight assignment based on the classification accuracy of each labeler in the training data. Let $q_i^{WA}$ be the number of correctly classified items out of the supervised items for labeler $i$, we define the weight to be
\begin{align}
\label{eqn_WA}
w^{WA}_i=\frac{q_i^{WA}}{\sum_{i=1}^L q_i^{WA}}.
\end{align}

\subsection{Exponential Weighted Algorithm (EWA)}
Exponential weighted algorithm sequentially adjusts weight of each labeler based on the loss of the predicted label and the true label \cite{Hastie01}.
For sake of convenience, let $x_{ij} \in \{0,1\}$ be the binary labels and $\ell(\hat{y}_j,y_j)=(\hat{y}_j-y_j)^2$ be the loss function. Let $w^{EWA}_{i,j}$ be the weight of labeler $i$ at stage $j$ with initial value $w^{EWA}_{i,0}=1/L$. The goal of EWA is to achieve low regret $R_{N_s}$, defined as
\begin{align}
\label{eqn_EWA_regret}
R_{N_s}=\sum_{j=1}^{N_s} \ell(\hat{y}_j,y_j)-\min_{i \in \{1,2,\ldots,L\}} \sum_{j=1}^{N_s} \ell(x_{ij},y_j),
\end{align}
the difference of loss between collective classification and the best labeler.
The predicted label for item $j$ is
\begin{align}
\label{eqn_EWA_predict}
\hat{y}_j=\textnormal{ceil} \left( \frac{\sum_{i=1}^L w^{EWA}_{i,j} x_{ij}}{\sum_{i=1}^L w^{EWA}_{i,j}} -\frac{1}{2} \right),
\end{align}
where $\textnormal{ceil}(z)$ is the ceiling function that accounts for the smallest integer that is not less than $z$.
The weight is updated according to
\begin{align}
\label{eqn_EWA_update}
w^{EWA}_{i,j+1}=w^{EWA}_{i,j} \exp\left(-\eta \ell(x_{ij},y_j)\right).
\end{align}
By setting $\eta=\sqrt{\frac{8 \ln L}{N_s}}$, it is proved in \cite{Littlestone94} that $R_{N_s} \leq \sqrt{\frac{N_s}{2} \ln L}$. That is, the regret to the best labeler scales with $O(\sqrt{N_s})$ and the average regret per training sample $\frac{R_{N_s}}{N_s}$ scales with $O(\frac{1}{\sqrt{N_s}})$.

\subsection{Multi-class Adaboost (MC-Ada)}
For multi-class Adaboost \cite{Zhu06}, each labeler acts as a weak classifier $f_i(\cdot)$, and in the training stage it finds the best labeler according to a specified error function.
In each round the weights of the supervised data are updated so that the algorithm can find a labeler with better classification capability for
the misclassified training samples. The weight of each classier is determined according to the error function and
the final classier is the weighted combination of every labeler.
The multi-class Adaboost algorithm proposed in \cite{Zhu06} is summarized as follows:
\begin{enumerate}
  \item Initialize all weights of the training data to be $\alpha_j=\frac{1}{N_S}$. For $i = 1,2,....,L$ repeat Step 2 to Step 5
  \item Find the best labeler that minimizes the error \\
$err_i = \sum\nolimits_{j=1}^{N_S} \frac{\alpha_j\mathds{1}_{\{y_j\neq f_i(X_j)\}}}{\sum\nolimits_{j=1}^{N_S} \alpha_j}$
  \item Set $w_i = \log\frac{1-err_i}{err_i} + \ln (K-1)$
  \item Set $\alpha_j \leftarrow \alpha_j \cdot \exp(w_i \mathds{1}_{\{y_j\neq f_i(X_j)\}})$ for $j=1,2,\ldots,N_s$
  \item Normalize $\alpha$ to have unit norm
  \item The final classifier is \\$f^{MC-Ada}(X_j)={\arg\max}_{k=\{1,2,\ldots,K\}}\sum\nolimits_{i=1}^{L} w_i \mathds{1}_{\{k = f_i(X_j)\}}$
\end{enumerate}

\subsection{Conventional Support Vector Machine (C-SVM)}

Given supervised data $\{X_j,y_j\}_{j=1}^{N_s}$, conventional SVM aims to solve the optimization problem \cite{Hastie01}
\begin{align}
\label{eqn_SVM_primal}
&\min_{w,b,\xi}~\frac{1}{2}  \|w\|_2^2 + \frac{C}{N_s}\sum_{j = 1}^{N_s} \xi_j \\
&\textnormal{subject~to}~y_j(w^T X_j-b) \geq 1 - \xi_j,~\xi_j \geq 0~\forall j \in\{1,2,\ldots,N_s\}, \nonumber
\end{align}
where $\xi_j \geq 0$ is the soft margin and $C$ is a tuning parameter. By representing (\ref{eqn_SVM_primal}) in dual form, it is equivalent to solving
\begin{align}
\label{eqn_SVM_dual}
 &\max_{\sigma} -\frac{1}{2}\sum_{j = 1}^{N_s} \sum_{z = 1}^{N_s} \sigma_j \sigma_z y_j y_z X_j^TX_z+ \sum_{j = 1}^{N_s} \sigma_j \\
&\textnormal{subject~to}~0 \leq \sigma_j \leq \frac{C}{N_s}~\forall j \in\{1,2,\ldots,N_s\}. \nonumber
\end{align}

Let $\sigma^*=[\sigma_1,\sigma_2,\ldots,\sigma_{N_s}]^T$ be the solution of (\ref{eqn_SVM_dual}), then the optimal weight and intercept in (\ref{eqn_SVM_primal}) are $w^*=\sum_{j=1}^{N_s} \sigma_{j}^* y_j X_j$ and $b^*$ is the average value of $y_j-{w^*}^T X_j$ for all $j$ such that $0<\sigma^*_j<\frac{C}{N_s}$.
Therefore the binary classifier is
\begin{align}
f^{C-SVM}(X_j)&=\textnormal{sign}\left( {w^*}^T X_j+b^*\right) \nonumber \\
&=\textnormal{sign}\left( \sum_{z=1}^{N_s} \sigma_z^* y_z X_z^T X_j+b^*\right).
\end{align}

\subsection{Multi-class Support Vector Machine (MC-SVM)}
In \cite{Crammer02}, a multi-class support vector machine approach is proposed by imposing a generalized hinge loss function as a convex surrogate loss function of the training error. Let $M=[M_1,M_2,\ldots,M_K]^T$ be the matrix containing all weighting vectors $M_k \in \mathds{R}^L$ for class $k$, the generalized hinge loss function is defined as $\max_{r \in \{1,2,\ldots,K\}} \{M_r^T X_j-1-\delta_{y_j,r}\}-M_k^T X_j$, where $\delta_{y_j,r}$ is the Kronecker delta function such that $\delta_{y_j,r}=1$ if $y_j=r$ and $\delta_{y_j,r}=0$ otherwise.

By introducing the concepts of soft margins and canonical form of separating hyperplanes, MC-SVM aims to solve the optimization problem
\begin{align}
\label{eqn_MC-SVM}
&\min_{M,\xi}~\frac{\lambda}{2}  \|M\|_2^2+\sum_{j=1}^{N_s} \xi_j \\
&\textnormal{subject to}~M_{y_j}^T X_j+ \delta_{y_j,k}-M_{k}^T X_j \geq 1-\xi_j~\forall j,k \nonumber,
\end{align}
where $\xi=[\xi_1,\xi_2,\ldots,\xi_{N_s}]^T$ is the vector of slack variables accounting for margins with $\xi_j \geq 0$, $\|M\|^2_2=\sum_{k,j} M_{kj}^2$
is defined as the $\ell_2$-norm of the vector represented by the concatenation of $M'$s rows, and $\lambda$ is the regularization coefficient.

Let $\mathbf{1}_z$ be a vector of zero entries except that its $z$th entry being $1$, $\mathbf{1}$ be the vector of all ones, and $\tau=[\tau_1,\tau_2,\ldots,\tau_{N_s}]$ be a $K$-by-$N_s$ matrix. The optimization problem in (\ref{eqn_MC-SVM}) can be solved in dual form by
\begin{align}
\label{eqn_MC-SVM_Dual}
&\max_{\tau}~-\frac{1}{2} \sum_{j=1}^{N_s} \sum_{z=1}^{N_s} (X_j^T X_z) \cdot (\tau_j^T \tau_z)+ \lambda \sum_{j=1}^{N_s} \tau_j^T e_{y_j}\\
&\textnormal{subject to}~\tau_j \leq \mathbf{1}_{y_j},~\mathbf{1}^T\tau_j=0~\forall j \nonumber.
\end{align}
Consequently the classifier for MC-SVM is
\begin{align}
f^{MC-SVM}(X_j)={\arg\max}_{k \in \{1,2,\ldots,K\}} \left\{ \sum_{z=1}^{N_s} \tau_{kz} X_z^T X_j \right\}.
\end{align}
For algorithmic implementation using multi-class SVM, $X_j$ is extended to a $K \times L$-by-$1$ vector, where the label for item $j$ provided by labeler $i$ is represented as $\mathbf{1}_{x_{ij}}$. For instance, the label $3$ of a $4$-class SVM is represented by $[0~0~1~0]^T$.

\subsection{EM Algorithm}
For sake of convenience, let $x_{ij} \in \{0,1\}$ be the binary labels. Following the definitions in \cite{Raykar10}, let $\alpha_i=P(x_{ij}=1|y_j=1)$ be the probability of correct classification of labeler $i$ when $y_j=1$ and $\beta_i=P(x_{ij}=0|y_j=0)$ be the probability of correct classification of labeler $i$ when $y_j=0$. Given the
crowdsourced data $X=\{X_1,X_2,\ldots,X_{N}\}$, the task is to estimate the parameters $\alpha=[\alpha_1,\alpha_2,\ldots,\alpha_L]^T$ and $\beta=[\beta_1,\beta_2,\ldots,\beta_L]^T$, where we denote the parameters by $\theta=\{\alpha,\beta\}$.

Assuming the items are independently sampled, the likelihood function of observing $X$ given $\theta$ is
\begin{align}
\label{eqn_EM_likelihood}
P(X|\theta)=\prod_j^{N} P(x_{1j},x_{2j},\ldots,x_{Lj}|\theta)=\prod_j^{N} P(X_j|\theta).
\end{align}
The maximization problem can be simplified as we apply EM algorithm \cite{Dempster77}, which is an efficient iterative procedure to compute the maximum-likelihood solution in presence of missing/hidden data. Here, we regard the unknown hidden true label $y_j$ as the missing data.
Define
\begin{align}
a_j&=\prod_{j=1}^N {\alpha_i}^{x_{ij}} (1-\alpha_i)^{1-x_{ij}};~
b_j=\prod_{j=1}^N {\beta_i}^{x_{ij}} (1-\beta_i)^{1-x_{ij}};    \\
u_j&=P(y_j=1|X_j,\theta) \propto P(X_j|y_j=1) \times PD(y_j=1|\theta)  \nonumber \\
&= \frac{a_j v}{a_j v+b_j(1-v)}
\end{align}
by the Bayes rule with $v=\frac{1}{N}\sum_{j=1}^{N} u_j$.
The complete loglikelihood can be written as
$\ln P(X,y|\theta)=\sum_{j=1}^N y_j \ln v a_j +(1-y_j) \ln (1-v) b_j$.
The EM algorithm is summarized as follows:\\
\textbf{E-step:}
Since $\mathds{E} [\ln P(X,y|\theta)] = \sum_{j=1}^N u_j \ln v a_j + (1-u_j) \ln (1-v) b_j$, where the expectation is with respect to $P(y|X,\theta)$, we
compute $v=\frac{1}{N}\sum_{j=1}^{N} u_j$ and update $u_j=\frac{a_j v}{a_j v b_j(1-v)}$. \\
\textbf{M-step:} Given the updated posterior probability $u_j$ and the observed data $X$, the parameters $\theta$ can be estimated by maximizing the conditional expectation of correct specification probability, i.e., $\alpha_i$ and $\beta_i$, by
\begin{align}
\label{eqn_EM_Mstep}
\alpha_i=\frac{\sum_{j=1}^N u_j x_{ij}}{\sum_{j=1}^N u_j};~\beta_i=\frac{\sum_{j=1}^N (1-u_j)(1-x_{ij})}{\sum_{j=1}^N (1-u_j)};
\end{align}

The binary classifier is built upon the converged posterior probability $u_j$, i.e., $f^{EM}(X_j)=\textnormal{ceil}\left(u_j-\frac{1}{2}\right)$.
For initial condition, the conventional (unsupervised) EM algorithm adopts $u_j=\frac{1}{L} \sum_{j=1}^N x_{ij}$.
Since supervised data are available, we also propose to modify the initial condition to be $u_j=\frac{1}{L} \sum_{j=1}^N w^{WA}_i x_{ij}$, where the weight $w^{WA}_i$ is defined in Sec. \ref{subsec_WA}. The experimental results show that the collective classification accuracy can be improved by
setting the initial condition as the weighted average of the supervised data.

\subsection{Naive Bayes (NB) Sampler/Classifier}

Naive Bayes sampler is a generative classifier that assumes conditional independence among components in $X_j$. The classifier can be represented
as $f^{NB}(X_j)={\arg\max}_{k=\{1,2,\ldots,K\}}\hat{\pi}_k\hat{g}_k(X_j)$, where $\hat{\pi}_k$ is the estimated prior of the training data, and $\hat{g}_k(X_j)$ is the estimated probability mass function $g_k(X_j)=\prod_{i=1}^{L} g_k^{(i)}(x_{ij})$, and
$g_k^{(i)}(x_{ij})$ is the marginal probability mass function of the random variable $X_{ij}\vert{Y=k}$.

The prior is estimated using the Dirichlet prior, which accounts for uniformly distributed prior. This alleviates the problem that data samples of some classes do not appear in the training data. We thus have the estimators $\hat{\pi}_k = \frac{\phi_k}{N_s+K}$ and $\hat{g}_k^{(j)}(z_l)=\frac{\phi_{kl}^{j}+1}{\phi_k+K}$, where $\phi_k=\vert{\{j: y_j=k\}}\vert$ and $\phi_{kl}^{(i)}=\vert{\{j: y_j=k \wedge x_{ij}=z_l\}}\vert$.\\

\section{The Proposed Supervised Collective Classification Method}
\label{sec_RIP}

We propose a saddle point algorithm for supervised collective classification, where the weight of each labeler is the solution of a convex optimization problem of the form
\begin{align}
\label{eqn_RIP_primal}
\min_{w}~T \left(w,\{X_j\}_{j=1}^{N_s},\{y_j\}_{j=1}^{N_s}\right)+\lambda R(w),
\end{align}
and $\lambda$ is the regularization parameter.
The function $T$ is a convex surrogate loss function associated with the training error  $\frac{1}{N_s}\sum_{j=1}^{N_s} \mathds{1}_{\{f(X_j) \neq y_j\}}$.
In particular, we consider hinge loss function  $h(z)=\max(0,1-z)$ and therefore $T=\frac{1}{N_s} \sum_{j=1}^{N_s} h(y_j w^T X_j)$.
The function $R(w)$ is a convex regularization function on weighting vector $w$. In this paper, we consider the $\ell_1$-norm regularization functions, i.e., $R(w)=\|w\|_1=\sum_{i=1}^L|w_i|$.
The $\ell_1$-norm regularization function favors the sparsity structure of the weighting vector $w$ and therefore it aims to assign more weights on the experts
(labelers with high classification accuracy) hidden in the crowds.

With the hinge loss function, the formulation in (\ref{eqn_RIP_primal}) can be rewritten as
\begin{align}
\label{eqn_RIP_margin}
&\min_{w}~\frac{1}{N_s} \sum_{j=1}^{N_s}\xi_j+\lambda R(w) \\
&\textnormal{subject~to}~y_j w^T X_j \geq 1-\xi_j,~\xi_j \geq 0, j=1,2,\ldots,N_s, \nonumber
\end{align}
where $\xi_j$ accounts for the soft margin of the classifier \cite{Hastie01}.

The Lagrangian of (\ref{eqn_RIP_margin}) is
\begin{align}
\label{eqn_Lagragian}
\mathfrak{L}(w,\xi,\alpha,\beta)&=\frac{1}{N_s} \sum_{j=1}^{N_s}\xi_j+\lambda R(w) \\
&~~~- \sum_{j=1}^{N_s} \alpha_j (y_j w^T X_j-1+\xi_j) - \sum_{j=1}^{N_s} \beta_j \xi_j,  \nonumber
\end{align}
where $\alpha_j,\beta_j \geq 0$ are the Lagrange multipliers.
The dual optimization problem of (\ref{eqn_RIP_margin}) becomes
\begin{align}
\max_{\alpha,\beta,\alpha_j,\beta_j \geq 0} \min_{w,\xi} \mathfrak{L}(w,\xi,\alpha,\beta).
\end{align}
Fixing $\alpha$, $\beta$, and $w$, the value $\xi$ that minimizes $\mathfrak{L}$ will satisfy the following equation:
\begin{align}
\label{eqn_xi}
\frac{\partial \mathfrak{L}}{\partial \xi_j} = \frac{1}{N_s} - \alpha_j - \beta_j=0.
\end{align}
Note that $(\ref{eqn_xi})$ implies $0 \leq \alpha_j, \beta_j \leq \frac{1}{N_s}$.
Substituting (\ref{eqn_xi}) to (\ref{eqn_Lagragian}), the Lagrangian can be simplified to
\begin{align}
\label{eqn_Lagragian_2}
\mathfrak{L}(w,\alpha)&=\lambda R(w)- \sum_{j=1}^{N_s} \alpha_j (y_j w^T X_j-1).
\end{align}
Therefore the dual optimization problem becomes
\begin{align}
\label{eqn_dual}
\max_{\alpha,~0 \leq \alpha_j \leq \frac{1}{N_s}} \min_{w} \mathfrak{L}(w,\alpha).
\end{align}
The solution to $(\ref{eqn_dual})$ is a saddle point of $\mathfrak{L}$ that can be obtained by iteratively solving the inner and outer optimization problem
and updating the corresponding parameters in (\ref{eqn_dual}) \cite{Boyd11}.

Since our regularization function $R(w)=\|w\|_1$ is not differentiable when $w_i=0$ for some $i$. We use the subgradient method \cite{Boyd04,Boyd11} to solve the
inner optimization problem. The subgradient $g$ of $\|w\|_1$ at a point $w_0$ has to satisfy $\|w\|_1 \geq \|w_0\|_1 + g^T(w-w_0)$ for all $w$.
Consider a one-dimensional regularizer function $|w|$. Since $|w|$ is everywhere differentiable except when $w=0$, substituting $w_0=0$ we have the constraint on the subgradient at $0$ that $g \leq \frac{|w|}{w}\in [-1,1]$. For $w \neq 0$, $g$ is the gradient of $|w|$ that $g=1$ if $w>0$ and $g=-1$ if $w<0$. Extending these results to $R(w)=\|w\|_1$, we define the (entrywise) projection operator of a $L$-dimensional function $g$ as $Proj_g(\theta)=[Proj_g(g_1),\ldots,Proj_g(g_L)]^T$, where
\begin{align}
Proj_g(g_i)=\left\{
                          \begin{array}{ll}
                            g_i, & \textnormal{if}~|g_i| \leq 1,  \\
                            \frac{g_i}{\|g\|_\infty}, & \textnormal{if}~|g_i| > 1,
                          \end{array}
                        \right.
\end{align}
and $\|g\|_\infty=\max_{i} g_i$ is the infinity norm of $g$.
Therefore the projection operator $Proj_g$ guarantees that the function $g$ to be a feasible subgradient of $\|w\|_1$.

Fixing $\alpha$, differentiating $\mathfrak{L}$ with respect to $w$ by using the subgradient $g$ as the gradient at the non-differentiable points gives
\begin{align}
g=\frac{1}{\lambda} \sum_{j=1}^{N_s} \alpha_j y_j X_j.
\end{align}
By the subgradient method the iterate of $w$ at stage $t+1$ given $\alpha^{(t)}$ is updated by
\begin{align}
w^{(t+1)}=w^{(t)} \pm s_w Proj_g \left(\frac{1}{\lambda} \sum_{j=1}^{N_s} \alpha^{(t)}_j y_j X_j \right),
\end{align}
where $s_w$ is the constant step length and we set $w^{(0)}=\mathbf{0}$, the vector of all zeros. The sign of the subgradient is determined so that $\mathfrak{L}(w^{(t+1)},\alpha^{(t)}) \leq \mathfrak{L}(w^{(t)},\alpha^{(t)})$.

Similarly, for the outer optimization problem, given $w^{(t+1)}$, the gradient of $\mathfrak{L}$ in (\ref{eqn_Lagragian_2})
with respect to $\alpha_j$ is $1-y_j {w^{(t+1)}}^T X_j$. Since $0 \leq \alpha_j \leq \frac{1}{N_s}$, define the (entrywise) projection operator of a $N_s$-dimensional function $\alpha$ as
\begin{align}
Proj_\alpha (\alpha_j)=\left\{
                          \begin{array}{ll}
                            \alpha_j, & \textnormal{if}~0 \leq |\alpha_j| \leq \frac{1}{N_s},  \\
                            \frac{\alpha_j}{\|\alpha\|_\infty N_s}, & \textnormal{if}~\alpha_j > \frac{1}{N_s},  \\
                            0, & \textnormal{if}~\alpha_j < 0.
                          \end{array}
                        \right.
\end{align}
The projection operator $Proj_\alpha$ projects $\alpha$ onto its feasible set.
The iterate of $\alpha$ at stage $t+1$ given $w^{(t)}$ is updated by
\begin{align}
\alpha^{(t+1)}= Proj_\alpha \left(\alpha^{(t)}+s_\alpha
   \textnormal{vec}\left(1-y_j {w^{(t+1)}}^T X_j\right) \right),
\end{align}
where $s_\alpha$ is the constant step length, and $\textnormal{vec}\left(1-y_j {w^{(t+1)}}^T X_j\right)=[1-y_1 {w^{(t+1)}}^T X_1,\ldots,1-y_{N_s} {w^{(t+1)}}^T X_{N_s}]^T$.
Since $\alpha$ relates to the vector of importance of the training samples,
we set $\alpha^{(0)}=\frac{1}{N_s} \mathbf{1}$ as the initial point, which means that all training samples are assumed to be equally important in the first place.
The algorithm keeps updating the parameters $\alpha$ and $w$ until they both converge. In this paper we set the convergence criterion to be the $\ell_2$ norm (Euclidean distance) between the old and newly updated parameters (e.g., the $\ell_2$ norm is less than $0.01$).
The proposed algorithm is summarized as follows:

\begin{algorithm}
\caption{The proposed supervised collective classification algorithm}
\label{algo_collective}
\begin{algorithmic}
\State \textbf{Input:} training samples ${\{X_j\}}_{j=1}^{N_s}$, training labels ${\{y_j\}}_{j=1}^{N_s}$, regularization parameter $\lambda$
\State \textbf{Output:} optimal weighting vector $w^*$
\State Initialization: $\alpha^{(0)}=\frac{1}{N_s} \mathbf{1}$, $w^{(0)}=\mathbf{0}$, $t=0$
\While{$\alpha^{(t)}$ and $w^{(t)}$ do not converge}
\State Compute $g=\frac{1}{\lambda} \sum_{j=1}^{N_s} \alpha_j^{(t)} y_j X_j$.
  \If{$\mathfrak{L}\left(w^{(t)} - s_w Proj_g(g),\alpha^{(t)} \right) \leq \mathfrak{L}\left(w^{(t)},\alpha^{(t)} \right)$}.
   \State $w^{(t+1)}=w^{(t)} - s_w Proj_g(g)$
  \Else
    \State $w^{(t+1)}=w^{(t)} + s_w Proj_g(g)$
  \EndIf
\State $\alpha^{(t+1)}= Proj_\alpha \left(\alpha^{(t)}+s_\alpha \textnormal{vec}\left(1-y_j {w^{(t+1)}}^T X_j\right) \right).$
\State $t=t+1$
\EndWhile
\State For robust algorithm, set $w^*_i=w^*_i \mathds{1}_{\{w^*_i>0\}}$
\end{algorithmic}
\end{algorithm}

Since (\ref{eqn_RIP_primal}) imposes no positivity constraint on the elements of the weighting vector $w$, some entries of $w$ can be negative, which implies that one should not trust the labels generated by labelers with negative weights for collective classification. However, in practice altering labeler's labels
might be too aggressive and resulting in non-robust classification for the test data.
One way to alleviate this situation is to truncate the weights by setting $\widetilde{w}_i=w_i$ if $w_i > 0$ and $\widetilde{w}_i=0$ if $w_i \leq 0$.
That is, the labels from reliable labelers are preserved, whereas the labels from unreliable labelers are discarded.
We refer to this approach as the proposed robust method. Note that the proposed saddle algorithm can be adjusted to different convex surrogate loss functions $T$ and convex
regularization function $R$ following the same methodology.

\section{Performance Evaluation}
We compare the crowdsourcing algorithms introduced in Sec. \ref{sec_existing_methods} with the proposed method in Sec. \ref{sec_RIP} on a canonical crowdsourced dataset
and the collected multiple-choice exam datasets.
The canonical dataset is the text relevance judgment dataset provided in Text REtrieval Conference (TREC) Crowdsourcing Track in $2011$ \cite{Trec11}, where labelers are asked to judge the relevance of paragraphs excerpted from a subset of articles with given topics. Each labeler then generates a binary label that is either ``relevant'' or ``irrelevant''. This dataset is a sparse dataset in the sense that in average each labeler only labels roughly $26$ articles out of $394$ articles in total. The exam datasets contains science exam with $40$ questions and math exam with $30$ questions. There are $4$ choices for each question and therefore this is a typical multi-class machine learning task. These datasets are quite dense in the sense that almost every student generates an answer for each question.

The oracle classifier to be compared is the performance of the best labeler in the crowds. All tuning parameters are determined by leave-one-out-cross-validation (LOOCV) approach swiping from $0$ to $200$ for the training data.
The classification accuracy are listed in Table \ref{table_result}, where the parentheses in the row of best labeler means the number of correctly specified items of the best labeler, and the classifier of the highest classification accuracy is marked by bolded face.

For the TREC$2011$ dataset, when $10$ percent of items ($40$ items) are used to train the classifier, majority votes leads to around $0.8$ classification rate.
The classification accuracy has notable improvement by using weighted averaging, conventional SVM, and supervised EM algorithm. Naive Bayes sampler has worse performance due to limited training samples. Note that the proposed robust algorithm outperforms others by assigning more weights to reliable labelers and
discarding labels from unreliable labeler.

The science dataset is perhaps the most challenging one since there are no perfect experts (i.e., labelers with classification accuracy $1$) in the crowds and most of students
do not provide accurate answers.
Despite its difficulties, our proposed method still outperforms others. Note that in this case the proposed non-robust and robust methods have the same classification accuracy
since this dataset is non-sparse and the accuracy of answers in the training data and test data are highly consistent.

The math dataset is a relatively easy task since the majority of students have correct answers.
Consequently unsupervised methods tend to have the same performance as the oracle classifier.
In case of limited training data size ($5$ training samples), some supervised methods such as naive Bayes sampler and multi-class SVM suffer performance degradation due to insufficient training samples, whereas the proposed method attains perfect classification.

\section{Conclusion}
This paper provides an overview of unsupervised and supervised algorithms for crowdsourcing and proposes a supervised collective classification method where the weights of
each labeler is determined via a saddle point algorithm. The proposed method is capable of distinguishing reliable labelers from unreliable ones to enhance the classification
accuracy with limited training samples. The results on a benchmark crowdsourced dataset and the exam datasets collected by the authors show that the proposed method outperforms
other algorithms. This suggests that supervised collective classification methods with limited training samples can be crucial for crowdsourcing and relevant applications.

\section*{Acknowledgement}
The first author would like to thank Tianpei Xie and Dejiao Zhang at the University of Michigan for useful discussions.

\begin{table*} \label{table_result}
\caption{Descriptions of crowdsourced datasets and classification accuracy.}
\centering
\begin{tabular}{l*{5}{|c}r}
Description / Dataset            & TREC2011 & TREC2011 & Science  & Math  & Math  \\
\hline
training data size ($N_s$) & 40 & 60 & 10 & 5 & 10 \\
test data size ($N-N_s$)  & 354 & 334 & 30 & 25 & 20 \\
number of labelers ($L$)  & 689 & 689 & 183 & 559 & 559 \\
\hline \\
Method / Classification accuracy          & TREC2011 & TREC2011 & Science  & Math  & Math \\
\hline
best labeler (oracle)                            & 1 (82) & 1 (84) & 0.7 (30) & 1 (25)& 1 (20) \\
majority vote                               & 0.7938 & 0.7964 & 0.4667   & \textbf{1}   & \textbf{1} \\
weighted averaging                         & 0.8305 & 0.8323 & 0.4667   & \textbf{1}   & \textbf{1} \\
exponential weighted algorithm            & 0.8051 & 0.8084 & 0.2667   & 0.36   & 0.4  \\
conventional SVM                                      & 0.8333 & \textbf{0.8413} & 0.5      & 0.96   & \textbf{1}  \\
multi-class SVM                                  & X      & X      & 0.4333   & 0.52   & 0.7  \\
unsupervised EM                                  & 0.7881 & 0.7784 & 0.5      & \textbf{1}   & \textbf{1}  \\
supervised EM                                    & 0.8277 & 0.8174 & 0.5      & \textbf{1}  & \textbf{1}  \\
naive Bayes sampler                              & 0.6921 & 0.6707 & \textbf{0.5333}   & 0.64  & \textbf{1}  \\
multi-class Adaboost  & 0.8051 & 0.7994 & 0.5167   & 0.8489 & 0.885\\
\hline
The proposed method & 0.8277  & 0.8323 & \textbf{0.5333}   & \textbf{1}  & \textbf{1} \\
The proposed robust method & \textbf{0.8446}  & \textbf{0.8413} & \textbf{0.5333}   & \textbf{1}  & \textbf{1} \\
\hline
\end{tabular}
\end{table*}

\bibliographystyle{IEEEtran}
\bibliography{IEEEabrv,crowdsourcing}

\end{document}